\documentclass{article}

\usepackage{PRIMEarxiv}
\usepackage[utf8]{inputenc} 
\usepackage[T1]{fontenc}    
\usepackage{hyperref}       
\usepackage{url}            
\usepackage{booktabs}       
\usepackage{amsfonts}       
\usepackage{nicefrac}       
\usepackage{microtype}      
\usepackage{lipsum}
\usepackage{fancyhdr}       
\usepackage{graphicx}       
\usepackage{amsmath}
\usepackage{color}

\graphicspath{{media/}}     

\usepackage{lineno}

\title{Investigating the Capabilities of Deep Learning for Processing and Interpreting One-Shot Multi-offset GPR Data: A Numerical Case Study for Lunar and Martian Environments}

\author{
  Iraklis Giannakis, \\ 
  University of Aberdeen, School of Geosciences \\
  Aberdeen, United Kingdom\\
  \texttt{iraklis.giannakis@abdn.ac.uk} 
\And
   Craig Warren \\
   Northumbria University \\
   Northumbria, United Kingdom \\
   \texttt{craig.warren@northumbria.ac.uk}
  \And
  Antonios Giannopoulos \\
  The University of Edinburgh \\
  Edinburgh, United Kingdom \\
  \texttt{a.giannopoulos@ed.ac.uk}
  \And
  Georgios Leontidis \\
  Interdisciplinary Centre for Data and AI, \\
  University of Aberdeen \\
  Aberdeen, United Kingdom \\
  \texttt{georgios.leontidis@abdn.ac.uk}
  \And
  Yan Su \\
  Chinese Academy of Sciences \\
  Beijing, China \\
  \texttt{suyan@nao.cas.cn}
  \And
  Feng Zhou \\
  China University of Geosciences (Wuhan) \\
  Wuhan, China \\
  \texttt{zhoufeng@cug.edu.cn}
  \And
  Javier Martin-Torres \\
  University of Aberdeen, School of Geosciences \\
  Aberdeen, United Kingdom \\
  \texttt{javier.martin-torres@abdn.ac.uk}
  \And
  Nectaria Diamanti \\ 
  Aristotle University of Thessaloniki, Faculty of Sciences, Department of Geophysics\\
  Thessaloniki, Greece \\
  \texttt{ndiamant@geo.auth.gr}
  }

\begin{document}
\maketitle

\begin{abstract}
Ground-penetrating radar (GPR) is a mature geophysical method that has gained increasing popularity in planetary science over the past decade. GPR has been utilised both for Lunar and Martian missions providing pivotal information regarding the near surface geology of Terrestrial planets. Within that context, numerous processing pipelines have been suggested to address the unique challenges present in planetary setups. These processing pipelines often require manual tuning resulting to ambiguous outputs open to non-unique interpretations. These pitfalls combined with the large number of planetary GPR data (kilometers in magnitude), highlight the necessity for automatic, objective and advanced processing and interpretation schemes. The current paper investigates the potential of deep learning for interpreting and processing GPR data. The one-shot multi-offset configuration is investigated via a coherent numerical case study, showcasing the potential of deep learning for A) reconstructing the dielectric distribution of the the near surface of Terrestrial planets, and B) filling missing or bad-quality traces. Special care was taken for the numerical data to be both realistic and challenging. Moreover, the generated synthetic data are properly labelled and made publicly available for training future data-driven pipelines and contributing towards developing pre-trained foundation models for GPR.

\end{abstract}


\section{Introduction}

In-situ ground-penetrating radar (GPR) was introduced in planetary science in 2013 as part of the scientific payloads of Yutu-1 the Lunar rover of the Chinese Chang'E-3 mission \cite{fang2014lunar}. Since then, GPR was part of the scientific payloads of the Yutu-2 rover from the Chang'E-4 mission \cite{Li:2020}, the lander of the Chang'E-5 \cite{Fang:2023} and Chang'E-6 missions, the rover Perseverance \cite{Hamran:2020} from Mars 2020,  and Zhurong rover from Tianwen-1 mission \cite{Zhang:2023}. Moreover, GPR is planned to be used in the future missions Chang'E-7 \cite{Zou:2020} and ExoMars \cite{Herve:2020}, both missions expected to take place before 2030. Consequently, GPR is one of the most mainstream scientific payloads in the new era of space exploration, and one of the few in-situ geophysical methods applied in planetary science.

GPR is a mature geophysical method with a wide range of applications, each with its own uniqueness and challenges \cite{Daniels:2005}. To that extent, bespoke processing, modelling and interpretation tools are needed for almost each unique GPR application. Regarding planetary GPR, numerous processing and interpretation methodologies have been suggested, from hyperbola fitting \cite{Dong:2020, Feng:2022, Giannakis:2023} and migration \cite{Giannakis:2023} to centroid frequency shift analysis \cite{Giannakis:2023, Ding:2020_frequency, Eide:2023}. These approaches provide non-unique interpretations and often require laborious and manual processing combined with high-level of domain knowledge in GPR and geophysics in general. 

The issues mentioned in the previous paragraph are not exclusive to planetary GPR. To address them, machine learning (ML) and data-driven approaches have been proposed in an attempt to develop objective and fully automatic interpretation frameworks primarily for Terrestrial applications. Fine-tuned foundation computer vision models and self-supervised learning have been used for automatic hyperbola fitting for non-destructive testing \cite{Huang:2024, Li:2022_ML}; convolutional neural networks and dense neural networks were suggested for removing background and ringing noise \cite{Sun:2022, Sun:2022_b, Patsia:2023, Patsia:2023_2}; random forest was used to estimate the diameters of rebars in concrete inspections \cite{Giannakis:2019_c}; convolutional auto-encoders were utilised for full waveform inversion (FWI) trained using synthetic data \cite{Leong:2021}; U-net was used to fill missing traces \cite{Dai:2023}; and unsupervised clustering was used to segment radagrams from Yutu-2 rover \cite{Giannakis:2024_c}.

In the current paper we want to investigate the potential of machine learning for processing and interpreting one-shot multi-offset GPR data collected in a planetary setup \cite{Giannakis:2022_AGU_Abstract}. In particular, we will conduct a coherent numerical study using realistic and challenging synthetic data focusing on two problems A) automatic full waveform inversion and B) filling missing or bad quality traces. The models will be tuned for frequencies around $60-100$ MHz, similar to the range of the low frequency systems employed in the planetary rovers Yutu-1, Yutu-2 and Zhurong. Moreover our investigation will focus on media with low electric permittivity and conductivity as expected in the Lunar \cite{Olhoeft:1975} and Martian subsurface \cite{Zhang:2023_properties}. 

\begin{figure}
    \centering
    \includegraphics[width=0.95\textwidth]{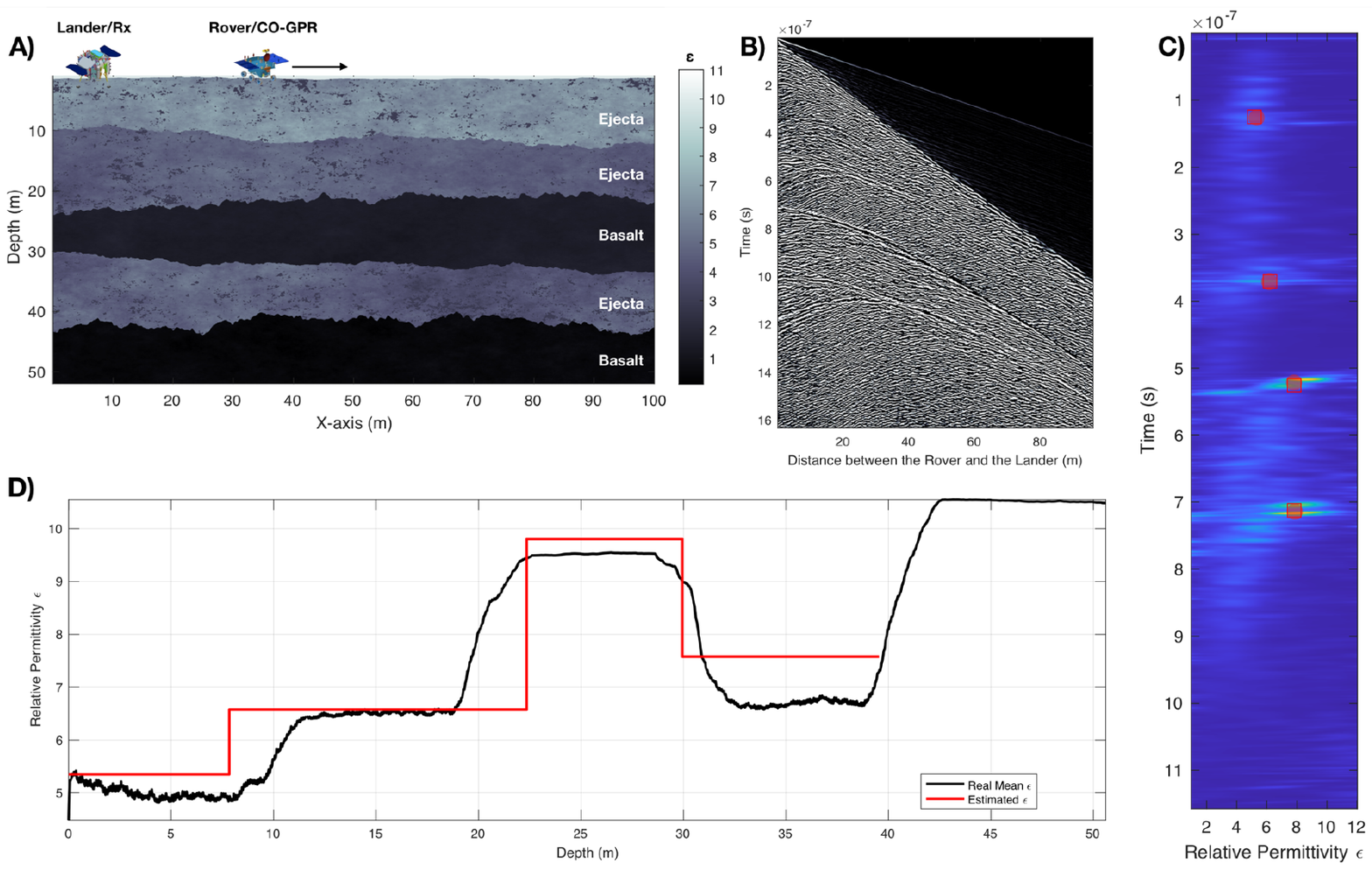}
    \caption{A) A generic one-shot multi-offset numerical case study. The discretisation of the model is 2 cm, and the central frequency of the source is 80 MHz. B) Processed B-Scan. C) NMO for picking velocities. D) The mean permittivity profile (black) and the reconstructed one (red) using Dix conversion \cite{Giannakis:2022_AGU_Abstract}. }
    \label{f1}
\end{figure}

\section{One-Shot Multi-Offset Configuration}

The most common measurement configuration in planetary in-situ GPR is the common-offset (CO), where the transmitter and receiver are moving along the measurement line while keeping their distance constant. CO has numerous drawbacks, one of the most important is the lack of a reliable methodology for calculating the subsurface permittivity. Hyperbola fitting (HF) is a very common approach used in CO-GPR for estimating the permittivity, however HF is reliable only for shallow layers \cite{Giannakis:2023}, since for deeper ones’ large boulders are necessary to give rise to visible hyperbolic reflectors; and the interpretation has a high rate of uncertainty with increasing depth \cite{Giannakis:2022_b}. 

 \begin{figure}
    \centering
    \includegraphics[width=0.75\textwidth]{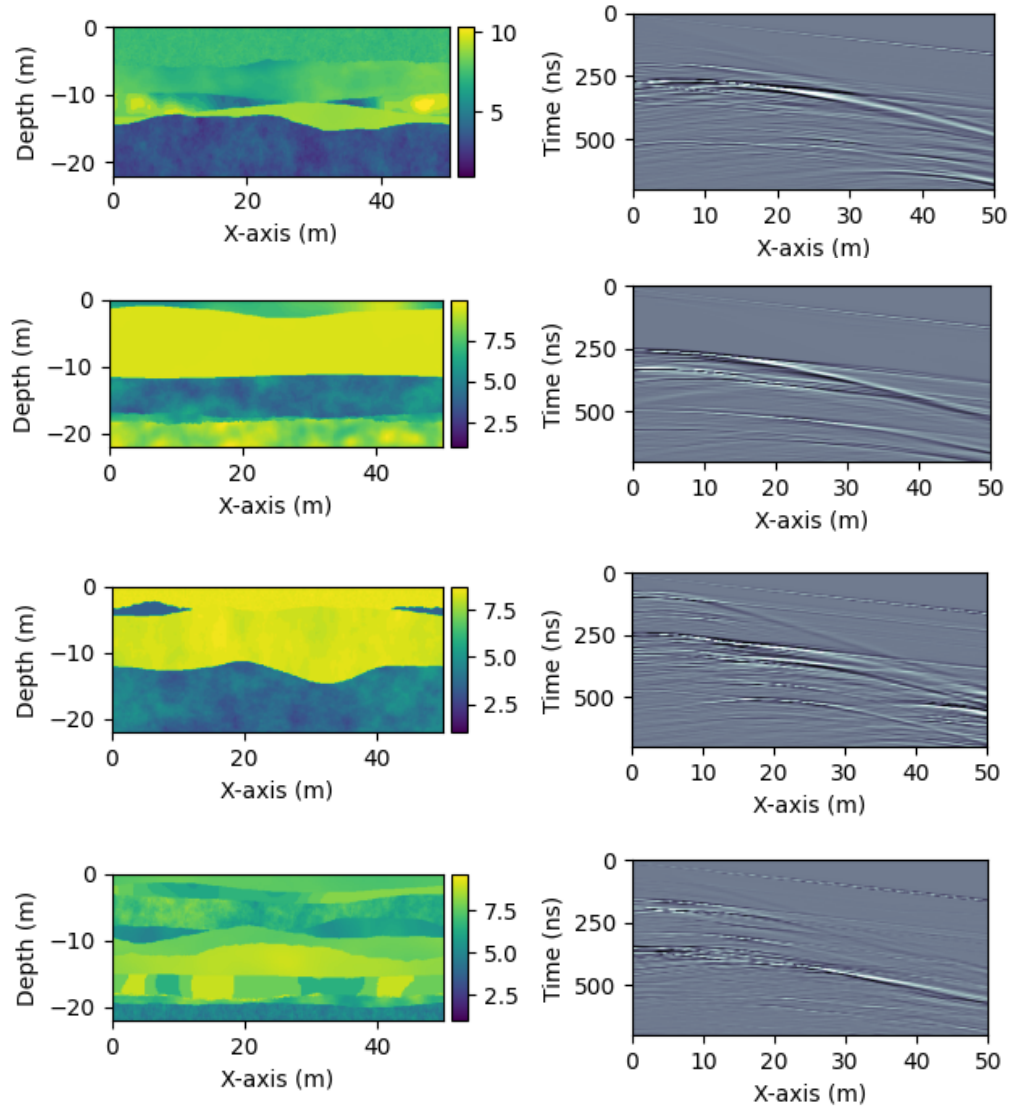}
    \caption{A set of training data from the 4400 samples used in this study. The colorbar illustrates the relative electric permittivity. Left figures depict the numerical models, and right figures are their corresponding B-Scans. A time-varying gain and signal saturation is applied to the B-Scans for illustration purposes. }
    \label{f2}
\end{figure}

The lack of an attainable method to calculate the permittivity in CO-GPR can be tackled using multi-offset (MO) configurations. where the receiver stands still while the transmitter is moving (or vice versa). MO-GPR has the ability to detect layers and accurately estimate their permittivity using the normal move-out (NMO) and Dix conversion \cite{Giannakis:2022_AGU_Abstract, Angelis:2022}. In a planetary setup, a MO-GPR can be realised --in theory-- by placing a receiver on the lander of the mission \cite{Giannakis:2022_AGU_Abstract}. The Chinese missions Chang’E-3, E-4 and Tianwen-1, all have landers, that could --in principle-- be utilised for MO-GPR. Figure \ref{f1} illustrates a numerical case study using one-shot MO-GPR for mapping Lunar ejecta. The processing applied was a typical NMO and Dix conversion \cite{Giannakis:2022_AGU_Abstract, Angelis:2022}. The discretisation of the model is 2 cm, and the central frequency of the source is 80 MHz. It is apparent that via this approach deep layers can be sufficiently detected and characterised.

 \begin{figure}
    \centering
    \includegraphics[width=0.95\textwidth]{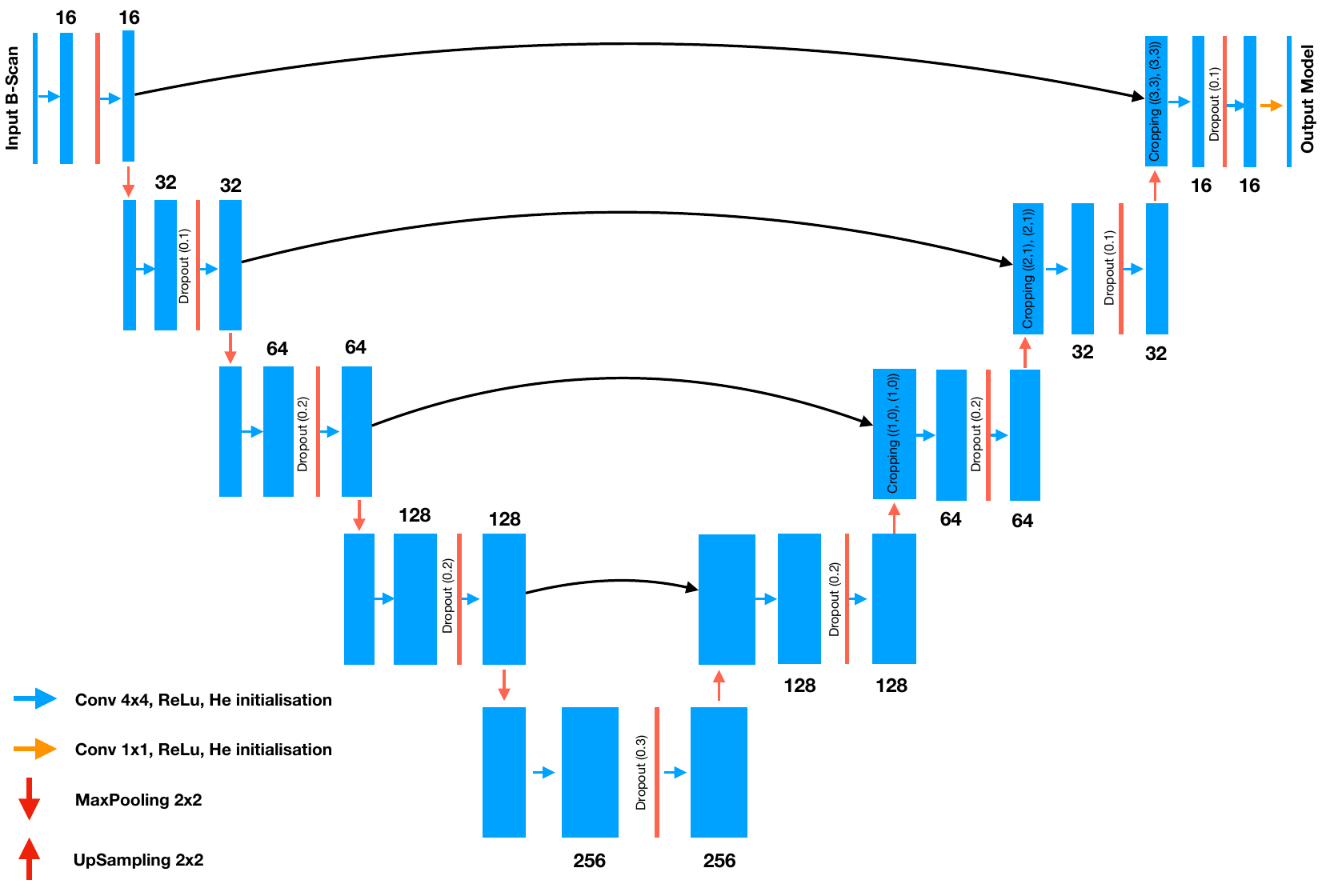}
    \caption{The architecture of the ensemble U-nets used for FWI. The input B-Scans have 230x230 dimensions each, and the output models have 224x224.}
    \label{f3}
\end{figure}

\section{Deep Learning - Processing and Interpretation}
In this section we will explore the capabilities of deep learning for processing and interpreting one-shot MO-GPR data. We will explore two problems A) Full-Waveform Inversion (FWI), and B) imputing missing traces. The section is divided into three subsections. The first subsection describes the synthetic training data used for training, providing details on how to generate them using the open-source electromagnetic simulator gprMax \cite{Warren:2016}. The second subsection suggests a deep learning scheme for automatic FWI using one-shot MO-GPR data. Finally, the last subsection investigates the capabilities of deep learning on reconstructing missing or bad quality traces.  

 \begin{figure}
    \centering
    \includegraphics[width=0.65\textwidth]{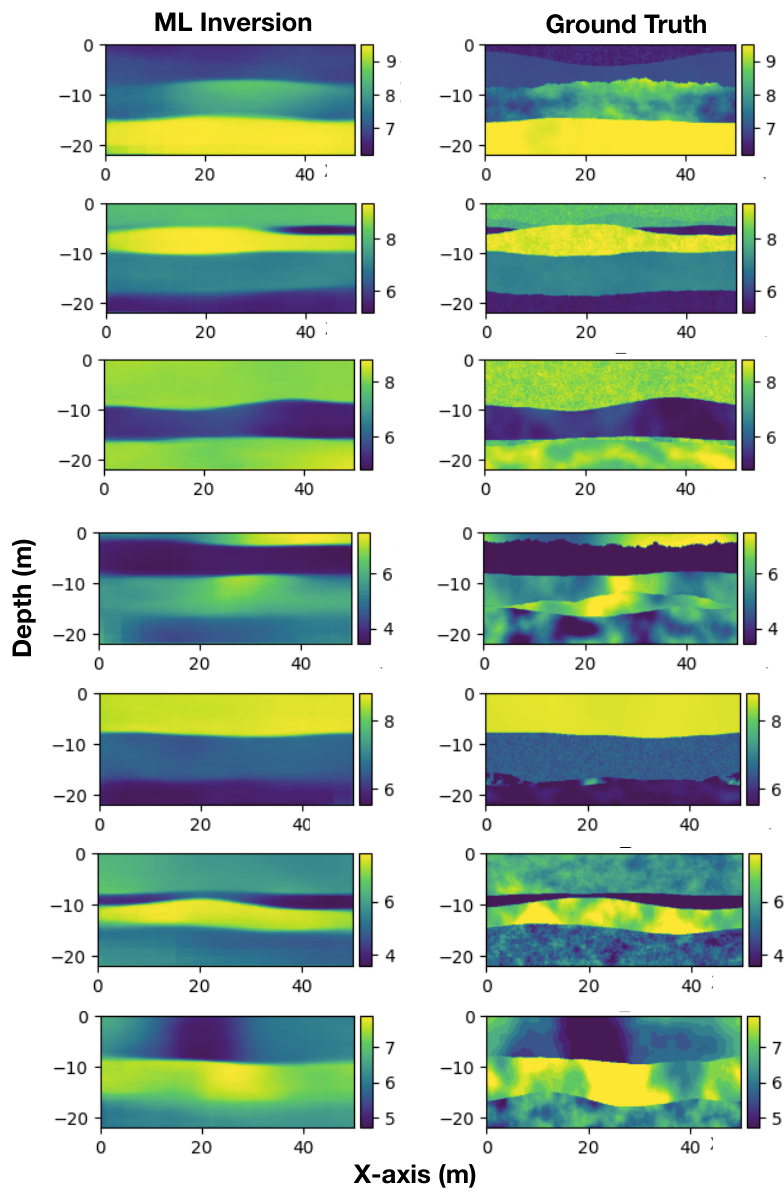}
    \caption{A set of examples comparing the ground truth permittivity models to the reconstructed ones using the suggested ensemble U-nets. These are unknown models that were not included in the training process. Colourbars depict the relative permittivity values. It is apparent that the proposed ensemble approach has the capability to reconstruct a smooth representation of the subsurface. }
    \label{f4}
\end{figure}

\subsection{Training Data}
The synthetic training data were generated using the open-source software gprMax \cite{Warren:2016}, an electromagnetic simulator that uses a second order accuracy (in both space and time) Finite-Differences Time-Domain (FDTD) method \cite{Taflove:2000}. GprMax is capable of running in parallel using multiple CPUs and GPUs \cite{Warren:2019} making it an appealing choice for generating big data. Moreover, gprMax is tuned for GPR modelling, and is also script-able allowing for the generation of randomly varying models. 

All the models of the training data are 2D with a uniform discretization step $\Delta x = \Delta y = 2$ cm. The time-step equals with $dt = 0.0467$ ns, which is 0.99 times the Courant limit. FDTD runs for 15000 iterations i.e. for a total of 700 ns. The dimensions of the model are $50.02 \times 26$ m, which results to a $2501\times 1300$ 2D grid. Perfectly Matched Layer (PML) is used to truncate the boundaries. In particular, a 60-layer thick PML is used in order to reduce the numerical artifacts from the surface waves on the free-space/soil interface \cite{Giannopoulos:2008}. The excitation source is placed 2 meters from the left boundary, and at 60-100 cm height from the surface. Both the shape and the central frequency of the pulse varies between training samples. The central frequency varies from 60-100 MHz, and the type of the pulse is randomly selected from \textit{gaussiandot, gaussiandotnorm, gaussiandotdot} and \textit{ricker} (see \href{https://docs.gprmax.com/en/latest/}{gprMax} documentation). Using different central frequencies and different types of pulses will allow for the model to generalise for unknown excitation sources. The polarisation of the pulse is orthogonal to the 2D domain. Along the surface, on the same height with the transmitter are 230 receivers, placed every 20 cm starting at a 2 meters distance from the excitation source.   

 \begin{figure}
    \centering
    \includegraphics[width=0.95\textwidth]{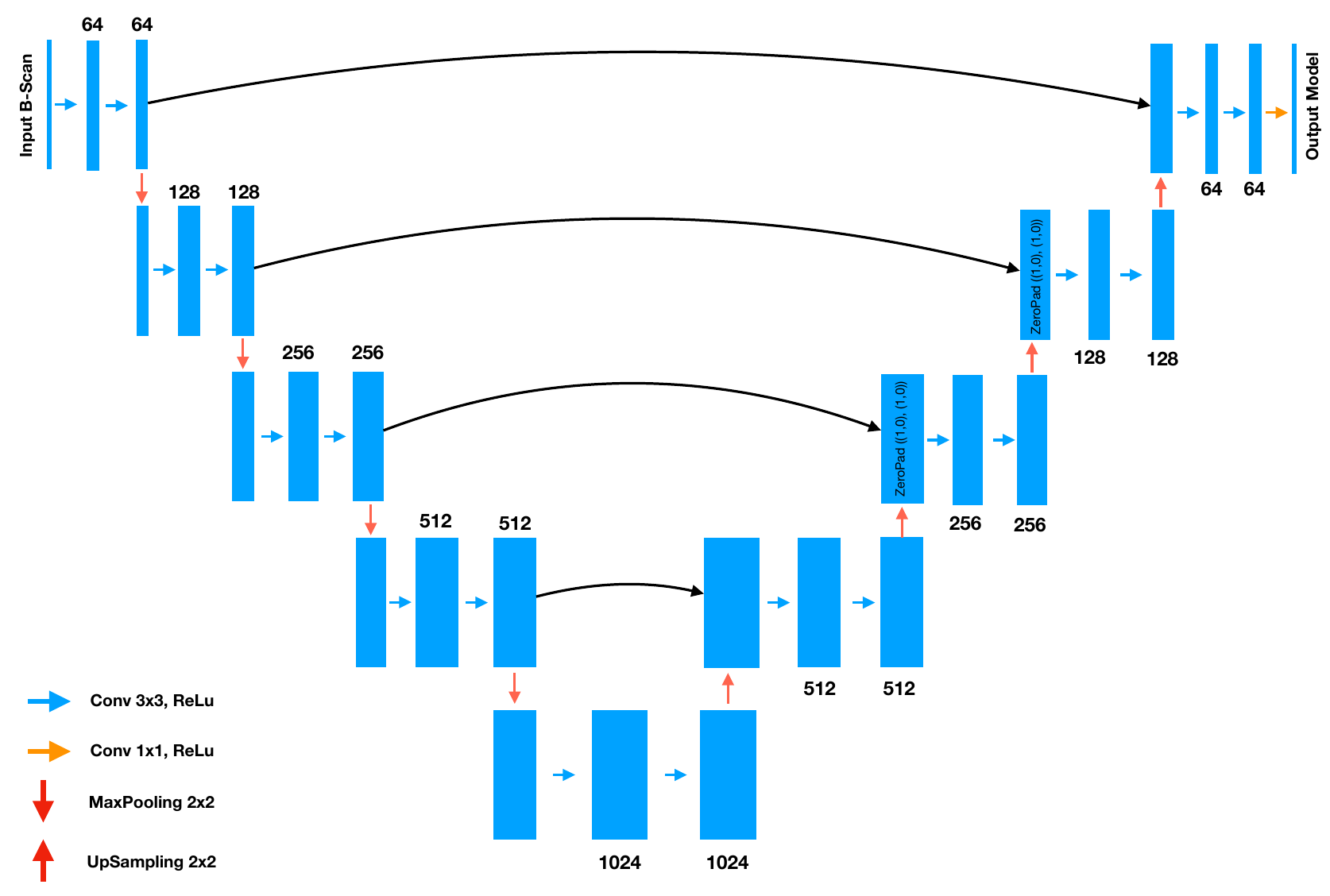}
    \caption{ The architecture of the U-net used for filling missing data. The inputs are the corrupted B-Scans each having 230x230 dimensions, and the outputs are the initial B-Scans.}
    \label{f5}
\end{figure}

Each of the models consists of a series of layers with random thickness, position, topography and dielectric properties. Moreover, within each layer the permittivity varies stochastically using fractal correlated noise with randomly selected fractal dimension \cite{Giannakis:2016}. The permittivity of the layers varies from $\epsilon = 2 - 10$, which is consistent with the permittivity values measured in the Lunar samples from the Apollo mission \cite{Olhoeft:1975}. The conductivity in planetary setups is negligible (due to the absence of liquid water), and is primarily related to oxides of titanium and iron (e.g. ilmenite) \cite{Strangway:1977}. For simplicity, we assume a correlation between permittivity and conductivity $\sigma = \epsilon/1000$, following the rationale that the abnormally high permittivity value of ilmenite and its dispersive properties will increase both the overall permittivity and the electromagnetic losses of an ilmenite-bearing formation \cite{Giannakis:2024}.

The training data consists of 4400 samples. Figure \ref{f2} shows 4 samples and their corresponding B-Scans. Varying layers, with different geometries and dielectric properties are included in the training data to increase the generalization capabilities of the resulting machine learning models. The initial raw B-Scan $\textbf{B}_{i}\in\mathbb{R}^{230\times 15000}$ is a matrix with 230 traces each of which having 15000 points length, where $i$ is the sample number. A time-varying gain is applied to every trace $A(n\cdot dt) = A(n\cdot dt)\cdot n^{3}$ to enhance late reflections, where $n$ is the time index. Subsequently each trace is compressed and resized to 230 dimensions, resulting to a new time step that is equal to 3.0434 ns. The processed and compressed B-Scan is now expressed as $\textbf{Q}_{i}\in\mathbb{R}^{230\times 230}$, a square matrix with 230 by 230 dimensions. 

The initial dielectric models can be expressed as matrices $\textbf{M}_{i}\in\mathbb{R}^{2501\times1300}$ containing the permittivity values of the  $50.02\times26$ m domain, with a spatial step of $\Delta x = \Delta y = 2$ cm. The first 4 meters of the model are free-space, and therefore are trimmed out i.e. $\textbf{M}_{i}\in\mathbb{R}^{2501\times 1100}$. Subsequently, a nearest neighbor interpolation is applied to reduce the dimensions of the data to $\textbf{W}_{i}\in\mathbb{R}^{224\times 224}$ resulting to the new spatial steps $\Delta x = 22.32$ cm and $\Delta y = 9.82$ cm. 

Conclusively, the final training data consist of 4400 compressed and processed (time-varying gain) B-Scans $\textbf{Q}_{i}\in\mathbb{R}^{230\times 230}$, and their corresponding dielectric models $\textbf{W}_{i}\in\mathbb{R}^{224\times 224}$.

 \begin{figure}
    \centering
    \includegraphics[width=0.65\textwidth]{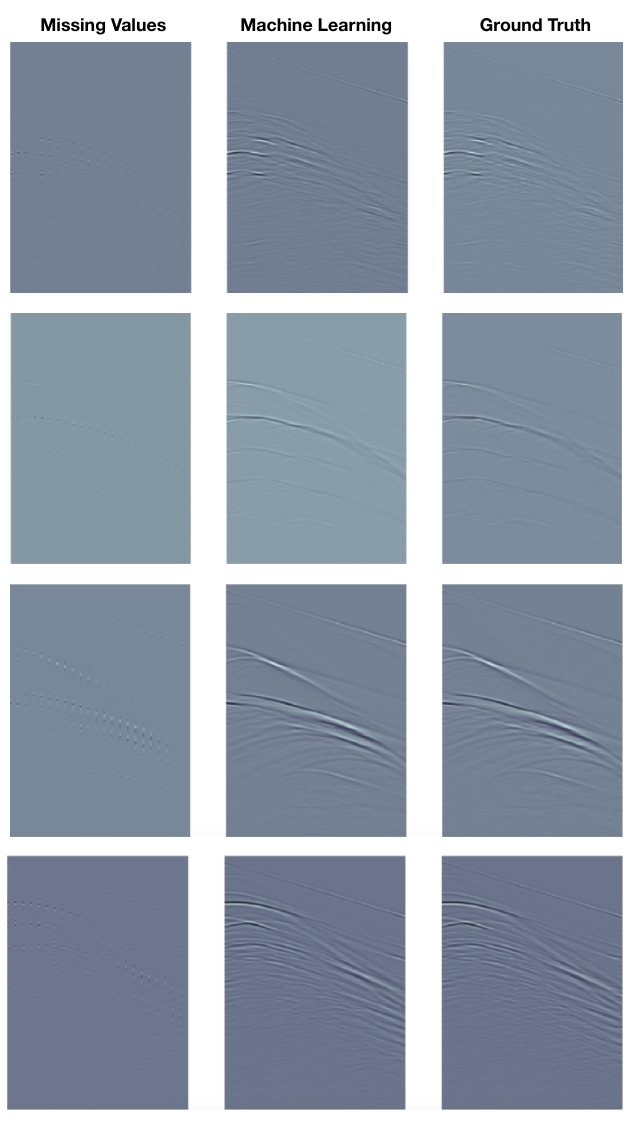}
    \caption{A set of testing examples used to evaluate the performance of U-net for reconstructing missing or bad quality MO-GPR data. These samples were not included in the training process. The label "Missing Values" correspond to the corrupted data with 9 traces removed every 2 meters. The label "Machine Learning" refers to the reconstructed B-Scans using the proposed U-net, and "Ground Truth" refers to the original B-Scans. It is apparent that deep learning is an appealing choice for treating missing values in one-shot MO GPR data.}
    \label{f6}
\end{figure}

\subsection{Full-Waveform Inversion}
In this section we aim to reconstruct the dielectric models $\textbf{W}_{i}$ from their corresponding B-Scans $\textbf{Q}_{i}$. In an abstract manner this is an image-to-image translation where the input image (B-Scan) $\textbf{W}_{i}\in\mathbb{R}^{230\times 230}$ is mapped into another image (dielectric model) $\textbf{Q}_{i}\in\mathbb{R}^{224\times 224}$. A very common deep learning methodology for this type of problems is U-net \cite{Ronneberger:2015}. More advanced methods suitable for image-to-image translation are Generative Adversarial Networks (GAN) \cite{Goodfellow:2014}, Pix2Pix \cite{Isola:2018}), Visual Transformers \cite{Dosovitskiy:2021} etc.. Despite the benefits of these techniques, in this paper we will employ a simple U-net in order to use it as a benchmark for comparing more advanced methodologies in the future. 

The architecture of the U-net is shown in Figure \ref{f3}. Dropout layers are used both in the encoding and decoding parts. The weights were initialised using the Kaiming He method \cite{He:2015}. The Adam optimizer \cite{Kinga:2017} was used to train the model. The learning rate was initially set to $5e-4$, after 55 epochs the learning rate reduced to $1e-4$, and after 30 epochs it was further reduced to $5e-5$ for another 30 epochs. The batch size equals with 5, and the loss function is the mean absolute difference. A validation set (10\%) is used to monitor the loss and stop training if the validation loss does not improve for 30 consecutive epochs. 

The U-net architecture is shown in Figure \ref{f3}. The training process described in the previous paragraph is executed 10 times using different initial weights, resulting to 10 different U-nets tuned for the synthetic training samples. The final model is an ensemble of the 10 U-nets. The performance of the model is evaluated in unknown scenarios that were not included in the training data. Figure \ref{f4} shows a set of examples of the FWI executed using the ensemble U-nets. It is apparent, that deep learning has the capability to reconstruct the dielectric properties of the subsurface even with just using one-shot MO-GPR data. More advanced approaches, combined with novel pre-processing techniques can further increase the overall accuracy, providing robust and reliable foundation models for FWI.

\subsection{Fill Missing Data}
Missing or bad quality data is a typical problem encountered both in Terrestrial and planetary applications of GPR \cite{Kumlu:2022, Zhang:2024_r}. In this section we explore the capabilities of deep learning on reconstructing missing traces in one-shot MO-GPR data. For the training we initially corrupt the data by removing traces from the original radagram. We keep one trace every 2 meters, i.e. for every 10 traces 9 traces are removed. The empty data are replaced with zeros. The corrupted data $\textbf{D}_{i}\in\mathbb{R}^{230\times230}$ are now used as inputs, and the initial data $\textbf{W}_{i}\in\mathbb{R}^{230\times 230}$ are the desired outputs. 

Similarly to the previous sub-section, U-net is used to map the causal relationship between $\textbf{D}$ and $\textbf{W}$. U-net is a very typical scheme used to fill missing data in geophysics, with promising results in MO seismic data \cite{Chai:2020}. The architecture of the employed U-net is shown in Figure \ref{f5}. The optimizer used is Adam with learning rate $1e-4$. The batch size is 16, and the loss function is the mean squared error. Similarly to the previous section 10\% of the data are used as a validation set with a patience of 30 epochs.

Figure \ref{f6} shows a series of examples demonstrating the capabilities of deep learning and U-net on reconstructing missing or bad quality data. Even in the absence of 90\% of the data, the trained U-net was capable of sufficiently reconstructing and predicting the missing values.  

\section{Data Availability - Kaggle Competition}
The training data used in this study are available in Kaggle in the community competition \textbf{GprMax Deep Learning Challenge 1 (GDLC-1)}. The competition aims at developing foundation deep learning models for FWI of one-shot MO-GPR data. Any submitted model must be trained subject to synthetic data (ideally using gprMax). The participants can either use the provided synthetic datasets, or they can use gprMax (or any software of their choice) to generate additional data to complement training. The participants can also apply any pre-processing they think necessary to the input files prior to training. The final evaluation metric is the mean absolute error between the ground truth permittivity of the testing models and the reconstructed ones using the proposed ML models.

\textbf{Files - Training Data}
\begin{itemize}
    \item 
\textbf{Training\_Bscan}: Folder that contains the .npy files for the processed (time-varying gain) B-Scans.
\item
\textbf{Bscan\_'n'.npy}: Input data. The nth Bscan saved in a numpy array, np.shape(B\_Scan\_'n'.npy) = (230,230).
\item
\textbf{Training\_Labels}: Folder that contains the .npy files for the ground truth i.e. relative permittivity structure.
\item
\textbf{Model\_'n'.npy}: Labels. The nth ground truth model saved in a numpy array, np.shape(Model\_'n'.npy) = (224,224).
\end{itemize}

\textbf{Files - Testing Data}

\begin{itemize}
\item
\textbf{Evaluation\_Dataset}: Folder that contains the .npy files for the processed (time-varying gain) B-Scans from the testing set.
\item
\textbf{Testing\_Bscan\_'n'.npy}: Input data. The nth testing Bscan saved in a numpy array, np.shape(B\_Scan'n'.npy) = (230,230).
\end{itemize}

The B-Scans in the training set are complete without missing traces, while the B-Scans in the evaluation set have some random missing set of traces. This is to add an extra challenge to the competition simulating missing data or bad quality data that need to be interpolated. The participants need to find a way to train their model for missing data or develop a way to fill the missing data prior to FWI.

\section{Conclusions}
A numerical case study was presented to demonstrate the potential capabilities of one-shot multi-offset measurement configuration combined with deep learning interpretation. Through a series of numerical tests we showcase that deep learning can sufficiently invert for the subsurface's' dielectric properties, and accurately reconstruct  missing or bad quality data. Despite trained with numerical data, these models can be used as foundation models for future transfer learning with real-data, and act as case studies for demonstrating the potential capabilities of different measurement configurations and machine learning models for planetary GPR.

\bibliographystyle{unsrt}  
\bibliography{references}

\end{document}